\documentclass[aps,prl,twocolumn,groupedaddress]{revtex4}

\usepackage {graphicx}

\begin{document}


\title{
The Band-Gap Problem in Semiconductors Revisited: \\
Effects of Core States and Many-Body Self-Consistency
}


\author{Wei Ku}
\altaffiliation[Present address: ]{Department of Physics, University of California, Davis, California 95616.}

\author{Adolfo G. Eguiluz}
\affiliation{
Department of Physics and Astronomy, The University of Tennessee, Knoxville, TN 37996-1200,\\
and Solid State Division, Oak Ridge National Laboratory, Oak Ridge, TN 37831--6030
}



\begin{abstract}
A novel picture of the quasiparticle (QP) gap in prototype semiconductors Si 
and Ge emerges from an analysis based on all-electron, self-consistent, \textit{GW} 
calculations. The deep-core electrons are shown to play a key role via the 
exchange diagram ---if this effect is neglected, Si becomes a semimetal. 
Contrary to current lore, the Ge 3$d$ semicore states (e.g., their 
polarization) have no impact on the \textit{GW} gap. Self-consistency improves the 
calculated gaps ---a first clear-cut success story for the Baym-Kadanoff 
method in the study of real-materials spectroscopy; it also has a 
significant impact on the QP lifetimes. Our results embody a new paradigm 
for \textit{ab initio} QP theory.
\end{abstract}


\maketitle


The modern ``band-gap problem'' originated with the realization that 
density-functional theory (DFT) \cite{Kohn}, implemented in 
the local-density approximation (LDA), failed drastically in the description 
of the fundamental excitation gap of semiconductors and insulators. A 
significant step forward was achieved in the mid-eighties, when the first 
\textit{ab initio} calculations of quasiparticle (QP) states were performed
\cite{Hybertsen,Godby} within Hedin's \textit{GW} 
approximation (GWA)
\cite{Hedin}. At the present time, it is 
nearly-universally accepted
\cite{Hybertsen,Godby,Hedin,Aulbur,Shirley,Rohlfing,Rojas,Hamada,Arnaud} that the GWA yields 
QP gaps in semiconductors and insulators to within 0.1 eV of experiment 
---which is the level of accuracy required in the study of transport in 
these materials \cite{Shirley}.

In this Letter we uncover a novel picture of the physical ingredients 
underlying the observed QP gap in Si and Ge. Central elements of this 
picture are the impact of the core electrons on the many-body problem for 
the states at the gap, and the role of self-consistency.
The \textit{GW} schemes alluded to above turn out to benefit from ``cancellation of errors''
involving the neglect of both effects.
Our results illustrate the practical importance of the Baym-Kadanoff conserving method
\cite{Baym} for the study of excitations in real materials.

These conclusions are arrived at by eliminating approximations which are 
routinely introduced in the implementation of Hedin's scheme. First, the 
usual \textit{GW} work invokes the pseudopotential (PS) approximation ---by which the 
core states are effectively eliminated from the gap problem. However, PS 
theory does not guarantee that a ``partitioning'' of the electrons into two 
groups may lead to an accurate description of the dynamical self-energy of the valence 
states ---which, according to DFT, is a non-linear functional of the \textit{total} 
density. Semicore states pose a special challenge
\cite{Shirley,Rohlfing}; significantly, on 
the basis of a phenomenological model, the indirect nature of the Ge gap has 
been assigned to an effect of the polarization of the 3$d$ states \cite{Shirley}.

Second, in most \textit{GW} calculations the Dyson equation (DEq) is not solved to 
self-consistency. However, it has been shown, for a Hubbard-type model, that 
this practice leads to a genuine violation of charge conservation
\cite{Schindlmayr}. Still, from the available self-consistent 
solutions of the DEq \cite{Groot95,Holm98,Schone} it has been 
inferred that, while self-consistency, at the \textit{GW} level, is a must in 
total-energy calculations \cite{Holm99}, the same is ``to be 
avoided'' in the study of spectroscopy \cite{Aulbur,Groot95,Holm98,Groot96}. 
This state of affairs is unsatisfactory, since self-consistency is a 
necessary condition for the fulfillment of all the conservation laws \cite{Baym} ---and, 
thus, for a proper theory of transport \cite{Faleev}.

In our \textit{GW} calculations all the electrons are taken into account in the 
evaluation of the valence-electron self-energy. Remarkably, the deep core 
states are found to play a significant role in the QP gap problem via the 
core-valence exchange diagram. An additional surprise is that the (presumably 
important) shallow Ge 3$d$ semicore states have no effect on the \textit{GW} gap. 
Self-consistency at the \textit{GW} level does improve the QP gaps; it also impacts the 
QP lifetimes. Our results for the Si gap and the indirect Ge gap al L 
agree with experiment very well. Other aspects of our calculated QP band 
structures provide signatures of physics beyond the GWA.

We recall that the exact self-energy is ``$\Phi $-derivable,''
\cite{Luttinger} $\Sigma \left[ G \right]\left( {1{1}'} \right) 
= \delta \Phi / \delta G\left( {{1}'1} \right)$, where $\Phi $ is the 
Luttinger-Ward ``free-energy'' functional; our notation stresses the fact 
that $\Sigma $ is a functional of the dressed Green's function $G$. Now, a 
$\Phi $-derivable $\Sigma $, obtained on the basis of an \textit{approximate} $\Phi 
$-functional, coupled with a self-consistent solution of the DEq, $G^{ - 
1}\left( {1,{1}'} \right) = G_0^{ - 1} \left( {1,{1}'} \right) - \left( 
{\Sigma \left[ G \right]\left( {1,{1}'} \right) - \left( {\mu - \mu _0 } 
\right)\delta \left( {1 - {1}'} \right)} \right)$, ensures that $G$ fulfills 
the conservation laws \textit{exactly} \cite{Baym}. [Here $1,{1}'$ denote 
space-time points; Matsubara times \textit{$\tau $} are defined for $0 \le \tau \le \beta 
\hbar $. $\mu $, the chemical potential for the correlated system with a 
fixed number of electrons, is obtained self-consistently with $G$; $\mu $ 
differs appreciably ($\sim$1eV) \cite{Ku2000} from its 
counterpart $\mu _0 $ for the reference one-electron system whose Green's 
function is $G_0 $.] The GWA \cite{Hedin} is defined by the 
$\Phi $-functional
\begin{equation}
\includegraphics[width=7.2cm,height=1.3cm]{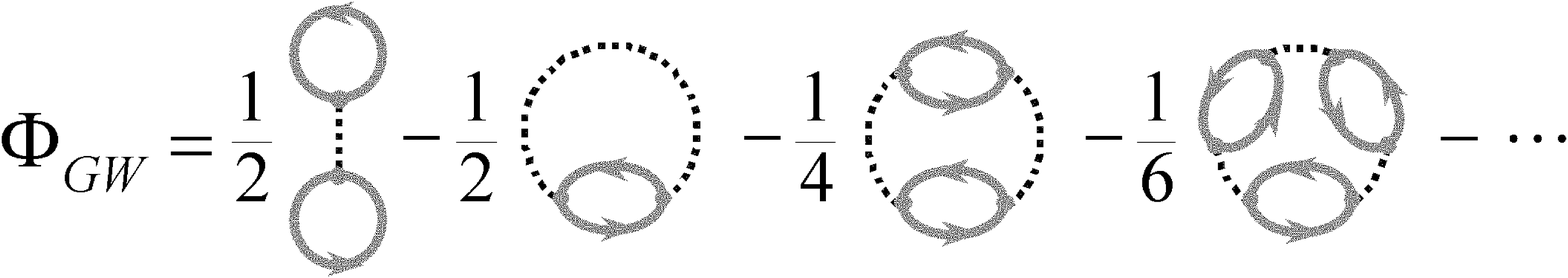},
\label{eq1}
\end{equation}
where the particle-hole bubbles are made up of $G$'s (not $G_0 $'s), and the 
dashes represent the Coulomb interaction $v$. Functional differentiation of 
$\Phi _{GW} $ yields
$\Sigma _{GW} \left[ G \right] = \Sigma _H \left[ G 
\right] + \Sigma _{xc} \left[ G \right]$, where $\Sigma _H \left[ G \right]$ 
is the Hartree term, and the exchange-correlation (XC) term is of the Hedin 
form $\Sigma _{xc} \left[ G \right]\left( {1,{1}'} \right) = - G\left( 
{1,{1}'} \right)W\left[ G \right]\left( {1,{1}'} \right)$, where $W\left[ G 
\right]\left( {1,{1}'} \right)$ is the screened interaction
\cite{Hedin}.

We work in the basis of the Kohn-Sham (KS) states
$\phi _{k,j} \left( x \right)$
provided by the full-potential, linearized augmented-plane-wave 
(FLAPW) method \cite{Blaha}; here $k$ is a wave vector in 
the Brillouin zone, and $j$ is a band index. Adopting (without lack of 
generality) the KS system as the reference one-particle system, described 
within the LDA, the DEq can be written as \cite{diag}
\begin{eqnarray}
 G_{k,j} \left( {\tau - {\tau }'} \right) = G_{k,j}^{LDA} \left( {\tau - 
 {\tau }'} \right) + G_{k,j}^{LDA}  \left( {\tau - \bar {\tau }_1 } \right) \nonumber\\
 \times \left[ {\Sigma _{k,j} \left( {\bar {\tau }_1 - \bar {\tau }_2 } \right) - 
 \left( {V_{k,j}^{LDA} + \left( {\mu - \mu _0 } \right)} \right)
 \delta \left( {\bar {\tau }_1 - \bar {\tau }_2 } \right)} \right] \nonumber \\
 \times G_{k,j} \left( {\bar {\tau }_2 - {\tau }'} \right),
 \label{eq2}
\end{eqnarray}
where $G_{k,j}^{LDA} \left( \tau \right) = - \frac{1}{\hbar }e^{ - 
\varepsilon _{k,j} \tau / \hbar }\left( {\theta \left( \tau \right) - n_F 
\left( {\varepsilon _{k,j} } \right)} \right)$, the KS eigenvalues 
$\varepsilon _{k,j} $ being measured from $\mu _0 $,
$V^{LDA}$ is the KS potential without the nuclear contribution,
and summation over variables with a bar on top is understood.

\begin{figure}[!tbp]
\includegraphics[width=3.00in,height=1.7in]{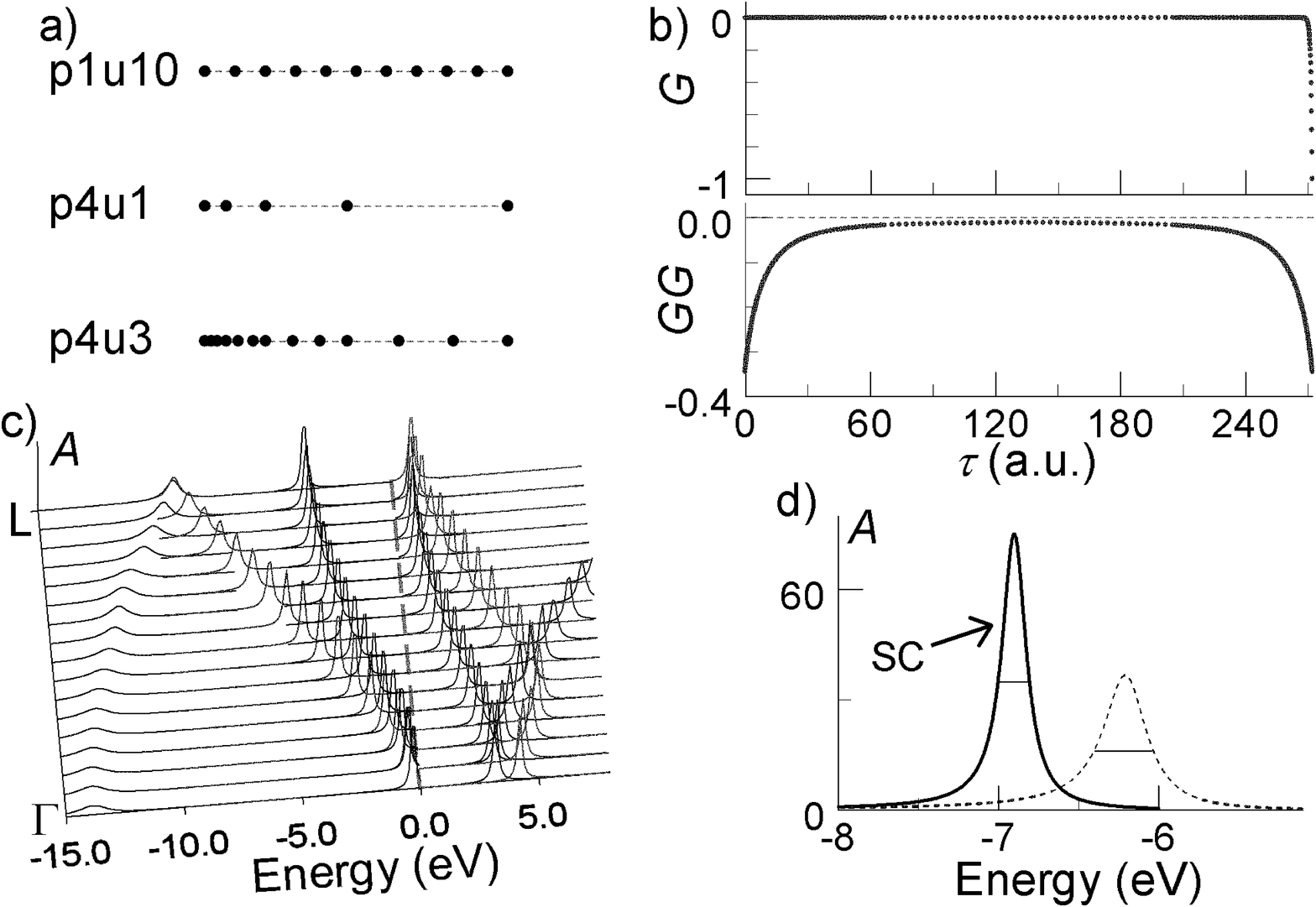}
\caption{
\label{fig1}
a): The PM used to solve the DEq on the \textit{$\tau $}-axis is defined by two 
integers: ``p'' is the ``order'' of the underlying non-uniform mesh, whose 
width doubles in each step; ``u'' is the number of uniform intervals into 
which the non-uniform intervals are partitioned. b): Typical \textit{$\tau $}-dependence of 
$G$ and of the particle-hole bubble (\textit{GG}); their exponential localization (and 
discontinuity) at $\tau = 0$ and $\beta \hbar $ is efficiently accounted for 
by our PM \cite{Ku2000}.
c): Spectral function $A_{k,j} \left( \omega \right)$ for Si along $\Gamma \mbox{L}$;
to aid visualization, small numerical broadening has been introduced.
d): $A_{k,j} \left( \omega \right)$ for the hole state at the midpoint of the occupied band;
the solid/dashed line denotes the self-consistent/first-iteration solution of Eq.~(\ref{eq2}).
}
\end{figure}

We solve Eq.~(\ref{eq2}) and the integral equation for $W\left[ G \right]\left( 
{1,{1}'} \right)$ (in the latter case, in reciprocal space) on the \textit{$\tau $}--axis. 
Our approach is ideally suited for a self-consistent evaluation of 
$G$ and should prove valuable in calculations beyond the GWA
\cite{Ku2000}. A novel feature of our scheme ---which is 
devoid of the cutoff effects encountered in \textit{$\omega $}-axis formulations
\cite{Schone}--- is the use of a non-uniform ``power mesh'' 
(PM) \cite{Ku2000} which, as outlined in Fig.~\ref{fig1}, accounts 
for the nature of both $G$ and the particle-hole bubble ---which are strongly peaked at 
the ends of the interval $0 \le \tau \le \beta \hbar $, being flat in 
between; moreover, the PM allows us to perform high-order interpolation 
(scaling $\sim$linearly) to generate all functions on the dense, uniform 
mesh required in the evaluation of the products entering the \textit{$\tau $}-integrals
\cite{Ku2000}. From the solution of Eq.~(\ref{eq2}) we evaluate the 
spectral function $A_{k,j} \left( \omega \right) = - \frac{1}{\pi }\text{Im}G_{k,j} 
\left( \omega \right)$ via analytic continuation of $G$ onto the real-\textit{$\omega $} axis
\cite{Ku99}. For the $j$-bands of interest, $A_{k,j} \left( 
\omega \right)$ shows a well-defined peak, whose $k - \omega $ dependence 
yields the QP band structure ---see Fig.~\ref{fig1}.

The convergence of our results was tested by varying the parameters involved 
in: \textbf{\textit{k}}-space sampling (we used 5x5x5 and 8x8x8 meshes), 
number of bands (14 and 24), number of (reciprocal lattice-) 
\textbf{\textit{K}}-vectors used in the evaluation of $\Sigma _{xc} $ (9, 
27, 51, and 65), temperature (4000, 2000, 1000, and 300 K), and PM 
mesh (p5u20, p6u10, and p6u20). The most demanding parameter, the number of 
\textbf{\textit{K}}-vectors kept in the valence exchange term, is associated with a 
monotonic opening of the Si gap.
(The core contribution converges even slower;
thus, a 6D real-space integral is evaluated instead.)
In the case of Si we estimate that the absolute gap is converged to $\sim$0.1 eV from below;
for Ge the convergence is even better, from both directions.

Table~\ref{tab1} provides the framework for a discussion of our results for the 
absolute QP gap (located at $\sim$80{\%} $\Gamma $X), the direct gap at 
$\Gamma $, and the occupied bandwidth of Si. Our results comprise two levels 
of implementations of the GWA: $(i)$ A self-consistent solution of Eq.~(\ref{eq2}), 
corresponding to the evaluation of the self-energy $\Sigma _{GW} \left[ 
{G_{GW} } \right]$ (3$^{rd}$ row); \textit{
this calculation represents a numerical realization of a conserving approximation} \cite{Baym}.
\textit{(ii) }A non-conserving calculation (4$^{th}$ row) corresponding to the use of the 
self-energy $\Sigma _{GW} \left[ {G_{LDA} } \right]$ ---the standard 
``$G_{LDA} \,W_{LDA} $'' approximation
\cite{Hybertsen,Godby,Hedin,Aulbur,Shirley,Rohlfing,Rojas,Hamada,Arnaud}. Clearly, the gaps 
obtained from $\Sigma _{GW} \left[ {G_{GW} } \right]$ are in good agreement 
with experiment \cite{Landolt}; the mechanisms behind this 
agreement turn out to be quite unexpected ---and instructive.

\begin{table}[!btp]
\caption{
\label{tab1}
QP band gaps and occupied bandwidth of Si.
Comparison of our \textit{all-electron} \textit{GW} results with experiment,
with the (approximate) all-electron \textit{GW} calculations of Refs. \cite{Hamada} and \cite{Arnaud},
and with a representative PS-based \textit{GW} calculation.
The 3$^{rd}$ row obtains from our fully conserving self-energy $\Sigma _{GW} \left[ {G_{GW} } \right]$;
the 4$^{th}$ row, and all rows below it, obtain from the non-conserving approximation
$\Sigma _{GW} \left[ {G_{LDA} } \right]$.
}
\begin{ruledtabular}
\begin{tabular}{lp{45pt}p{45pt}p{45pt}}
& Absolute gap& Direct gap at $\Gamma $& Occupied bandwidth \\
\hline
Experiment \cite{Landolt}& 1.17& 3.40& 12.5 $\pm $ 0.6 \\
LDA (FLAPW)& 0.52& 2.53& 12.22 \\
$\Sigma _{GW} \left[ {G_{GW} } \right]$& 1.03& 3.48& 13.53 \\
$\Sigma _{GW} \left[ {G_{LDA} } \right]$& 0.85& 3.12& 12.15 \\
$\sim$all-electron \cite{Hamada}& 1.01& 3.30& 12.21 \\
$\sim$all-electron \cite{Arnaud}& 1.00& 3.15& \ldots  \\
PS-based \cite{Hybertsen}& 1.29& 3.35& 12.04 \\
\end{tabular}
\end{ruledtabular}
\end{table}

Indeed, on the basis of several numerical tests, we uncovered, first of all, 
the role of the deep-core electrons. In one set of calculations we 
suppressed their contribution to the exchange self-energy ---2$^{nd}$ term 
in Eq.~(\ref{eq1})--- for the states at the gap \cite{exchange}. At 
the $\Sigma _{GW} \left[ {G_{LDA} } \right]$ level, the absolute gap ($\sim$0.85 eV,
see Table~\ref{tab1}) is then reduced by 0.9 eV
---\textit{i.e., the gap is closed, the ensuing QP band structure of Si corresponding to a semi-metal}.
While the size of this 
effect is surprising, its physics is easy to visualize: $(i) $The core electrons 
shield the attractive field of the nuclei (via the 1$^{st}$ term in Eq.~(\ref{eq1}),
thereby raising the energy of the valence and conduction states; \textit{(ii) }the 
exchange process partially compensates for this effect;
\textit{(iii)} the states across the gap have different amplitudes in the core region;
these amplitudes control the strength of the exchange. It is the larger 
lowering of the energy of the QP states below the gap, relative to those 
above it, \textit{due to the non-local core-valence exchange process}, that leads to this novel all-electron effect
\cite{core}.

Further insight into the role of the core electrons is obtained by 
simulating their contribution to $\Sigma _{xc} \left[ {G_{LDA} } \right]$ as 
$\Sigma _{xc}^{\mbox{from core}} \approx V_{xc}^{LDA} \left[ {n_{tot} } 
\right] - V_{xc}^{LDA} \left[ {n_{val} } \right]$, where $n_{tot} $ and 
$n_{val} $ are the total and valence densities, and $V_{xc}^{LDA} $ is the 
XC contribution to $V^{LDA}$. This uncontrolled ``LDA recipe'' yields a 
\textit{spurious} 0.15 eV additional opening of the Si gap \cite{Ku2000}. The 
significance of this test is that it accounts (Table~\ref{tab1}) for the difference 
between the \textit{approximate} FLAPW-based $\Sigma _{GW} \left[ {G_{LDA} } \right]$-level 
result of Hamada, Hwang, and Freeman \cite{Hamada} (1.01 eV) 
and our own (0.85 eV) ---Ref. \cite{Hamada} relies on this 
LDA \textit{ansatz }for the core contribution to the valence-electron self-energy, as the 
PS-based \textit{GW} schemes implicitly do.

From Table~\ref{tab1} we draw a second key message: contrary to current wisdom
\cite{Hedin,Groot95,Holm98,Schone,Groot96}, 
self-consistency does improve the quality of the calculated \textit{GW} gaps of Si; cf. 
rows 1, 3, and 4. Indeed, the additional opening of the gaps obtained from 
$\Sigma _{GW} \left[ {G_{GW} } \right]$ brings them closer to their 
experimental values. This effect is traced to the dressing of both $W$ and $G$. 
Indeed, the dynamical screening built into $W$, which greatly reduces the 
exaggerated Hartree-Fock (HF) gap, is weakened for the dressed $W$;
the dressing of $G$ widens the gap as well,
a trend most easily visualized within self-consistent HF \cite{Ku2000}.
We stress that\textit{ the success of our conserving scheme is intimately
related to the fact that we have carried out a full all-electron calculation}
\cite{exaggeration}.

Interestingly, self-consistency also has a significant effect on the QP 
\textit{lifetimes}, given by the inverse of the full-width at half-maximum (FWHM) of the QP 
peak in $A_{\vec {k},j} \left( \omega \right)$. As seen in Fig.~\ref{fig1}d) for the 
hole state at the midpoint of the Si band, the FWHM (solid line) is 
significantly reduced, relative to its non-self-consistent counterpart 
(dashes). This effect ---which has been ignored in the rapidly-growing 
literature on ``hot carrier'' lifetimes \cite{Echenique}--- 
recognizes two sources. First, fully dressed QP's scatter less frequently 
off the Fermi sea than in the $\Sigma _{GW} \left[ {G_{LDA} } \right]$ case. 
Second, in the latter, non-self-consistent case, the gap edges recognized by 
$G_{LDA}$ and Im$\Sigma _{GW} \left[ {G_{LDA} } \right]$ are different.
This mismatch in the gap edge, $\Delta $, exaggerates the FWHM according 
to $\sim(\varepsilon + \Delta )^2 - \varepsilon ^2 = 2\Delta \cdot 
\varepsilon + \Delta ^2$, where $\varepsilon $ is the QP energy measured 
from each gap edge, and $\Delta \sim$[$\left( {\mu - \mu _0 } \right)$ 
plus the shift of the respective edge relative to LDA] \cite{Ku2000}.
(N.B.: the QP states obtained from $\Sigma _{GW} \left[ {G_{LDA} } \right]$
have an unphysical finite lifetime, $\sim\Delta^{2}$, at the gap edges!)

From Table~\ref{tab1} it also follows that PS-based \textit{GW} schemes carry a built-in error 
\textit{which is comparable with the LDA gap }---cf. the representative PS-based result (1.29 eV
\cite{Hybertsen}; 7$^{th}$ row) with our corresponding 
all-electron result (0.85 eV; 4$^{th}$ row). In addition to the non-local 
many-body effect of the core elucidated above, there is the impact on the 
matrix elements of $\Sigma _{GW} $ of the removal of the oscillations of the 
valence wave functions at the atomic sites \cite{Ku2000}. 
Interestingly, the projector-augmented-wave result of Arnaud and Alouani
\cite{Arnaud} (6$^{th}$ row) provides an independent measure 
of the error due to the latter source. We stress that, at the $\Sigma _{GW} 
\left[ {G_{LDA} } \right]$ level, the uncontrolled PS approximation
\cite{Hybertsen,Godby,Hedin,Aulbur,Shirley,Rohlfing,Rojas} is masked by the 
neglect of self-consistency ---which introduces an opposite-sign error, and 
results in apparent agreement with the experimental gap \cite{exaggeration}.

\begin{table}[!b]
\caption{
\label{tab2}
QP band gap and occupied bandwidth of Ge.
In the 5$^{th}$ row we exclude the contribution from the 3$d$ states to the valence-state self-energy.
For other conventions, see Table~\ref{tab1}; for the 6$^{th}$ row, see text.
}
\begin{ruledtabular}
\begin{tabular}{p{65pt}p{39pt}p{39pt}p{39pt}p{45pt}}
& absolute gap: $\Gamma $L& Direct gap at $\Gamma $& Indirect gap: $\Gamma $X& Occupied Bandwidth \\
\hline
Experiment \cite{Landolt}& 0.74& 0.89& \ldots & \ldots  \\
LDA (FLAPW)& 0.35& -0.20& 0.66& 12.82 \\
$\Sigma _{GW} \left[ {G_{GW} } \right]$& 0.79& 1.51& 0.71& 14.77 \\
$\Sigma _{GW} \left[ {G_{LDA} } \right]$& 0.51& 1.11& 0.49& 13.12 \\
$\Sigma _{GW} \left[ {G_{LDA} } \right]$, no 3$d$'s& 0.51& 1.11& 0.49& 13.12 \\
PS-based +CPP \cite{Shirley}& 0.73& 0.85& 1.09& \ldots  \\
\end{tabular}
\end{ruledtabular}
\end{table}

Ge is isoelectronic with Si, and has the same diamond crystal structure. It 
is then reassuring that an analysis along the above lines confirms that our 
conclusions concerning the impact of the deep core states and many-body 
self-consistency apply in the case of Ge as well. Note, in particular, the 
1$^{st}$ column of Table~\ref{tab2}:
The additional opening of the \textit{indirect} gap at L 
obtained from $\Sigma _{GW} \left[ {G_{GW} } \right]$ (3$^{rd}$ row) 
---relative to the value obtained from $\Sigma _{GW} \left[ {G_{LDA} } 
\right]$; 4$^{th}$ row--- leads to excellent agreement with experiment
\cite{Landolt}.

\begin{figure}[!btp]
\includegraphics[width=2.99in]{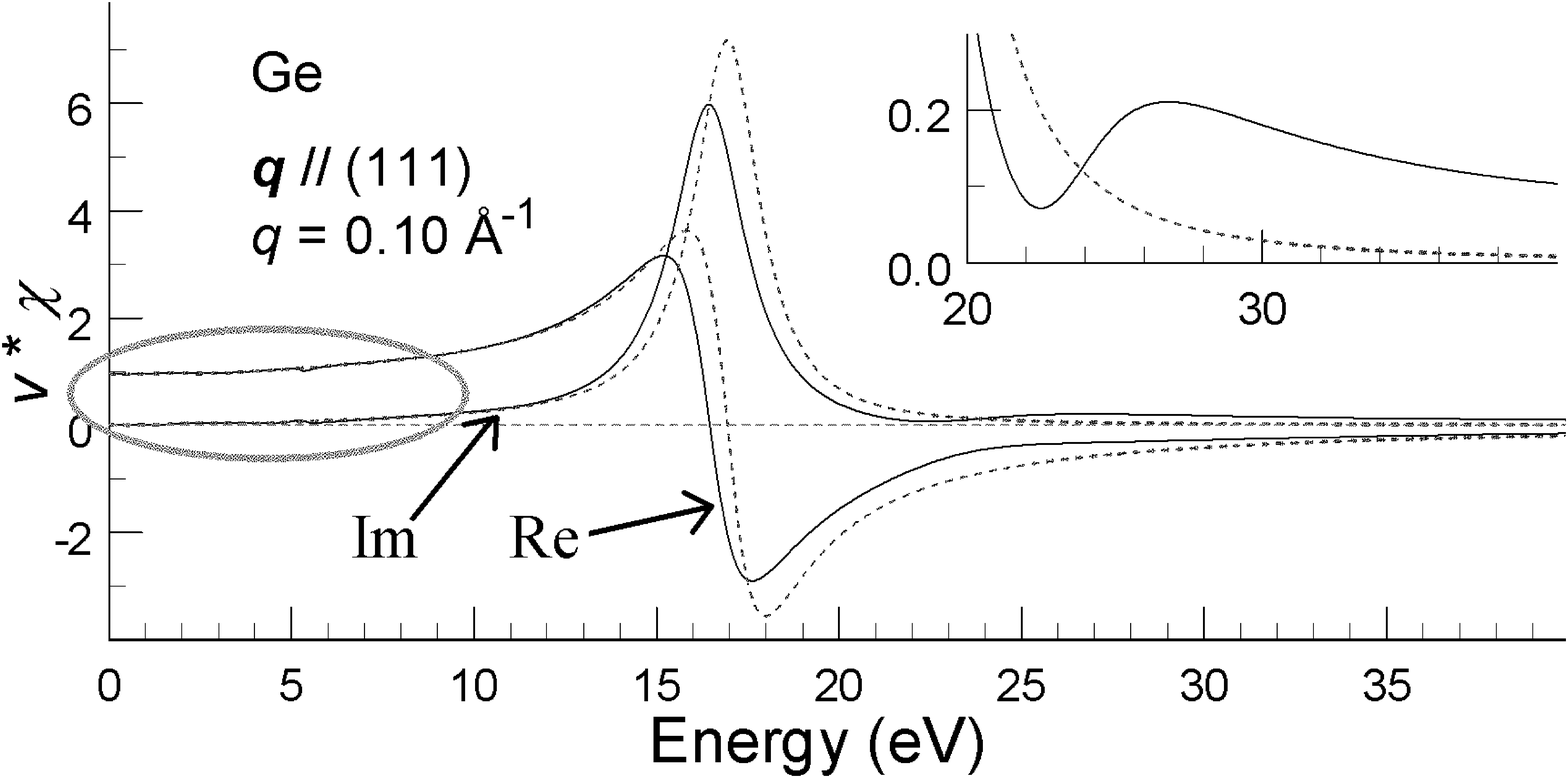}
\caption{
\label{fig2}
Solid lines: Real and imaginary parts of the density-response 
function of Ge, obtained in an adiabatic, local, implementation of 
time-dependent density-functional theory \cite{Ku99}. The 
dashes correspond to the elimination of transitions from the 3$d$ states. The 
3$d$-onset is highlighted in the inset.
}
\end{figure}

A significant issue in Ge is the role of the 3$d$ semicore states
(which lie $\sim$25eV below the gap). In fact, in the standard PS-based \textit{GW} 
approach, the correct topology of the QP band structure along $\Gamma $L 
\textit{---which is automatically produced in our results---} is obtained only upon introducing the 3$d$'s in the gap problem via a 
phenomenological core-polarization potential (CPP) model
\cite{Shirley}. Our all-electron approach places us in a 
position to address this issue. Shown in Fig.~\ref{fig2} is the density-response 
function \cite{Ku99} of Ge for a small wave vector, 
evaluated with and without the contribution from the 3$d$'s. Evidently, the 
screening is virtually identical in both calculations for energies up to 
$\sim$10 eV, by virtue of the weak transition strength of the 3$d$'s. Thus, 
the impact of the CPP model \cite{Shirley} is not physically 
justified. Moreover, if the 3$d$'s are excluded from \textit{both} $W$ and $G$, our all-electron 
QP band structure remains unchanged (5$^{th}$ row). We conclude that \textit{the 3d's play no role within the GWA}.

Now, although our calculated indirect gap at L agrees with experiment very well, other 
aspects of our \textit{GW} results for Ge are less satisfactory. As illustrated in 
Table~\ref{tab2} (3$^{rd}$ column), the empty states near X lie slightly below 
the lowest empty state at L ---contrary to experiment
\cite{Landolt}. This ordering, which is reversed in the 
absence of screening ---i.e., if $\Sigma $ is evaluated within HF
\cite{Ku2000}--- provides a signature of the limitations of 
the GWA (note also our results for the Ge direct gap and the Si and Ge bandwidths).
Thus, we expect that the localized Ge 3$d$ states 
may induce short-range correlation effects in $\Sigma $; within the 
conserving method, such effects are to be included by adding an appropriate 
$\Phi $-functional to Eq.~(\ref{eq1}) \cite{Eguiluz}.

In summary, our results embody a new paradigm for \textit{ab initio} QP theory, as we have 
demonstrated the non-trivial role played by the deep core states, and 
many-body self-consistency, in the QP gap problem; self-consistency ---and 
thus, the fulfillment of the conservation laws--- was also shown to impact 
the QP lifetimes. These effects, whose inclusion leads to excellent 
agreement with experiment for the Si absolute gap, and for the Ge indirect 
gap al L, are masked in the standard \textit{GW} schemes. The Ge 3$d$ semicore states were 
found to play no role in the \textit{GW} gap; their impact is likely to come via 
effects beyond the GWA.
Our results indicate that a fundamental description of the entire valence QP band structure
of Si and Ge within 0.1eV requires the inclusion of mechanisms beyond the GWA.

\begin{acknowledgments}
This work was supported by NSF and NERSC.
\end{acknowledgments}


\begin{references}

\bibitem{Kohn}
W. Kohn and L. J. Sham, Phys. Rev. \textbf{140}, A1133 (1965).

\bibitem{Hybertsen}
M. Hybertsen and S. G. Louie, \textbf{55}, 1418 (1985).

\bibitem{Godby}
R. W. Godby, M. Schl\"{u}ter, and L. J. Sham, Phys. Rev. Lett. \textbf{56}, 
2415 (1986).

\bibitem{Hedin}
L. Hedin, Phys. Rev. \textbf{139}, A796 (1965). 

\bibitem{Aulbur}
For a recent review, see, e.g., W. G. Aulbur, L. J\"{o}nsson, and J. W. 
Wilkins, in \textit{Solid State Physics, }edited by H. Ehrenreich (Academic, Orlando, 1999) Vol. 54, p.1.

\bibitem{Shirley}
E. L. Shirley, X. Zhu, and S. G. Louie, Phys. Rev. Lett. \textbf{69}, 2955 
(1992); Phys. Rev. B \textbf{56}, 6648 (1997).

\bibitem{Rohlfing}
M. Rohlfing, P. Kr\"{u}ger, and J. Pollmann, Phys. Rev. B \textbf{57}, 6485 
(1998).

\bibitem{Rojas}
H. N. Rojas, R. W. Godby and R. J. Needs, Phys. Rev. Lett. \textbf{74}, 1827 
(1995).

\bibitem{Hamada}
N. Hamada, M. Hwang, and A. J. Freeman, Phys. Rev. B \textbf{41}, 3620 
(1990).

\bibitem{Arnaud}
B. Arnaud and M. Alouani, Phys. Rev. B \textbf{62}, 4464 (2000).

\bibitem{Baym}
G. Baym and L. P. Kadanoff, Phys. Rev. \textbf{124}, 287 (1961); G. Baym, 
Phys. Rev. \textbf{127}, 1391 (1962).

\bibitem{Schindlmayr}
A. Schindlmayr, Phys. Rev. B. \textbf{56}, 3528 (1997).

\bibitem{Faleev}
S. V. Faleev and M. I. Stockman, Phys. Rev. B 62, 16707 (2000).

\bibitem{Groot95}
H. J. de Groot, P. A. Bobbert, and W. van Haeringen, Phys. Rev. B 
\textbf{52}, 11000 (1995).

\bibitem{Holm98}
B. Holm and U. von Barth, Phys. Rev. B \textbf{57}, 2108 (1998).

\bibitem{Schone}
W.-D. Sch\"{o}ne and A. G. Eguiluz, Phys. Rev. Lett. \textbf{81}, 1662 
(1998).

\bibitem{Holm99}
B. Holm, Phys. Rev. Lett. \textbf{83}, 788 (1999).

\bibitem{Groot96}
H. J. de Groot \textit{et al.}, Phys. Rev. B \textbf{54}, 2374 (1996); P. 
Garc\'{\i}a-Gonz\'{a}lez and R. W. Godby, \textit{ibid.} \textbf{63}, 075112 (2000).

\bibitem{Luttinger}
J. M. Luttinger, and J. C. Ward, Phys. Rev. \textbf{118}, 1417 (1960).

\bibitem{Ku2000}
Wei Ku, Ph. D. Thesis, The University of Tennessee (2000); Wei Ku, and A. G. 
Eguiluz, to be published.

\bibitem{Blaha}
P. Blaha, K. Schwarz, and J. Luitz, \textit{WIEN97} (Techn. Universit\"{a}t Wien, Austria, 
1999).

\bibitem{diag}
Here the ``$j$-diagonal'' approximation is employed as usual.

\bibitem{Ku99}
Wei Ku and A. G. Eguiluz, Phys. Rev. Lett. \textbf{82}, 2350 (1999).

\bibitem{Landolt}
\textit{Zahlenwerte und Funktionen aus Naturwissenschaften und Technik}, Landolt-Bornstein, New Series, Vol. III, Pt. 17a (Springer, New York, 
1982).

\bibitem{exchange}
The corresponding contribution of the deep core states via the higher-order 
terms in Eq.~(\ref{eq1}) is negligible.

\bibitem{core}
The relativistic 2P ($\kappa = - 2$, +1) states contribute more than 90{\%} 
to the opening of the Si gap, as they extend more in space, and have similar 
angular momentum character as the states that define the gap \cite{Ku2000}.

\bibitem{exaggeration}
The exaggerated opening of the gap obtained in a previous self-consistent 
\textit{GW} calculation \cite{Schone} is a consequence of the use of 
the PS approximation \cite{Ku2000}.

\bibitem{Echenique}
P. M. Echenique \textit{et al}., Chem. Phys. \textbf{251}, 1 (2000).

\bibitem{Eguiluz}
For a pedagogical discussion of the conservation laws,
see A. G. Eguiluz and Wei Ku, in \textit{Electron Correlations and Materials Properties},
edited by A. Gonis, N. Kioussis, and M. Ciftan (Kluwer-Plenum, N. Y., 1999), p. 329.

\end{references}

\newpage
\printtables
\newpage
\printfigures

\end{document}